\documentclass[10pt,aps,prl,twocolumn,superscriptaddress,reprint,showpacs,balance,notitlepage]{revtex4-1}

\usepackage{amssymb,amsmath,mathtools}
\usepackage{graphicx}
\usepackage{color}
\usepackage{hyperref,tikz,pgfplots}
\usepackage{listings}
\usepackage{orcidlink}
\usepackage{balance}
\usepackage{soul}
\graphicspath{{figures/}}

\makeatletter
\newcommand*{\balancecolsandclearpage}{%
  \close@column@grid
  \clearpage
  \twocolumngrid
}
\makeatother

\begin{document}

\definecolor{dkgreen}{rgb}{0,0.6,0}
\definecolor{gray}{rgb}{0.5,0.5,0.5}
\definecolor{mauve}{rgb}{0.58,0,0.82}

\newcommand{\todo}[1]{\vspace{0.2cm}\noindent \textbf{\textcolor{red}{#1}}}

\definecolor{mygreen}{rgb}{0.14, 0.84, 0.11}

\lstset{frame=tb,
  	language=Matlab,
  	aboveskip=3mm,
  	belowskip=3mm,
  	showstringspaces=false,
  	columns=flexible,
  	basicstyle={\small\ttfamily},
  	numbers=none,
  	numberstyle=\tiny\color{gray},
 	keywordstyle=\color{blue},
	commentstyle=\color{dkgreen},
  	stringstyle=\color{mauve},
  	breaklines=true,
  	breakatwhitespace=true
  	tabsize=3
}

\title{Fluid-induced snap-through instability of spherical shells}



\author{Pier Giuseppe Ledda\,\orcidlink{0000-0003-4435-8613}}
\affiliation{
Department of Civil, Environmental Engineering and Architecture, University of Cagliari, Via Marengo 2, 09123 Cagliari, Italy
}%

\author{Hemanshul Garg\,\orcidlink{0000-0002-0252-5877}}
\affiliation{Department of Mechanical and Production Engineering, {\AA}rhus University, Katrinebjergvej 89F, 8200 {\AA}rhus N, Denmark}

\author{Vitus \O{}stergaard-Clausen}
\affiliation{Department of Mechanical and Production Engineering, {\AA}rhus University, Katrinebjergvej 89F, 8200 {\AA}rhus N, Denmark}

\author{Lucas Krumenacker Rudzki}
\affiliation{Department of Mechanical and Production Engineering, {\AA}rhus University, Katrinebjergvej 89F, 8200 {\AA}rhus N, Denmark}

\author{Ahmad Madary}
\affiliation{Department of Mechanical and Production Engineering, {\AA}rhus University, Katrinebjergvej 89F, 8200 {\AA}rhus N, Denmark}

\author{Matteo Pezzulla\,\orcidlink{0000-0002-3165-8011}}
\email{matt@mpe.au.dk}
\affiliation{Department of Mechanical and Production Engineering, {\AA}rhus University, Katrinebjergvej 89F, 8200 {\AA}rhus N, Denmark}

\date{\today}

\begin{abstract}
\
\\
We study the snapping instability of a spherical elastic shell induced by a viscous flow, the umbrella flipping problem when life is at low Reynolds numbers. We combine precision desktop-scale experiments, fluid-structure simulations, shell theory, fluid mechanics, and scaling analysis to determine the instability threshold as a function of the geometrical and material parameters of the system. Building on these findings, we devise a snapping-based valve that passively and abruptly alters the hydraulic resistance of a channel, offering robust flow control without active components. Beyond the application, our study presents what we believe to be a prototypical example of fluid-induced elastic instability in viscous flow, providing a foundation for future explorations in soft hydraulics and flow-responsive structures.


\
\\
\\

\end{abstract}

\maketitle


Mechanical instabilities, such as Euler buckling of columns, are among the biggest concerns in structural design, but can also be tamed to enable new functionalities in soft matter~\cite{Forterre2005,Bertoldi2010,Holmes2013,Jawed2015,reis_perspective_2015}. Inspired by this, elastic instabilities have started to be tamed also in soft hydraulics~\cite{Christov2021}, where the compliance of soft channels and its soft components is being harnessed for improved hydraulic functionality~\cite{Leslie2009,Gomez2017,Box2020,Boyko2020,Peretz2020,Christov2021,garg2024passive}, towards plant-like systems that offer minimal complexity and decentralized control \cite{Aylmore1984}. In this paradigm, achieving robust, nonlinear, and often passive control of flow enables functions such as self-regulation and memory storage, without the need for electronics or active components \cite{Martinez2024,jensen2016sap,Louf2020}. While many of these research efforts have led to a variety of passive valves that give rise to nonlinear pressure-flow relationships, the case of flow-induced snapping, which leads to abrupt changes in shape and possibly hydraulic resistances \cite{Gomez2017}, have received far less attention, apart from studies on the snapping of thin sheets at high Reynolds numbers \cite{Goncharuk2023,Oshri2024}. In solid mechanics, a prototypical example of a three-dimensional continuous system undergoing such transition is that of spherical shells under pressure, a long studied problem in engineering~\cite{Zoelly1915,Hutchinson1967}. Not only has this problem proven to be fundamental for the aerospace industry, but it has also served as a foundation for more explorations in solid mechanics, where the existence of more than one stable state in a structure is seen as an opportunity instead of a mere route to failure~\cite{hutchinson2016buckling,Pezzulla2018,yan2021magneto}.


Here, we explore one such canonical scenario: the snap-through of a spherical elastic shell induced by viscous flow, the umbrella flipping problem when life is at low Reynolds numbers~\cite{Purcell1977}. This minimal system serves as a model to study a fluid-induced snapping instability and forms the foundation for designing passive valves that operate through geometric reconfiguration. Via scaling analysis, we first focus on singling out the key dimensionless parameters at play, and we then proceed to a more detailed analysis of our problem via desktop-scale experiments, shell and fluid mechanics modeling, and fluid-structure interaction (FSI) simulations. We determine the critical instability threshold as a function of the material and geometrical parameters of the system, and then proceed to the design of a snapping-based valve, which offers nonlinear passive hydraulic control.

\begin{figure}[t]
\vspace{0.2cm}
\centering
\includegraphics[scale=1.02]{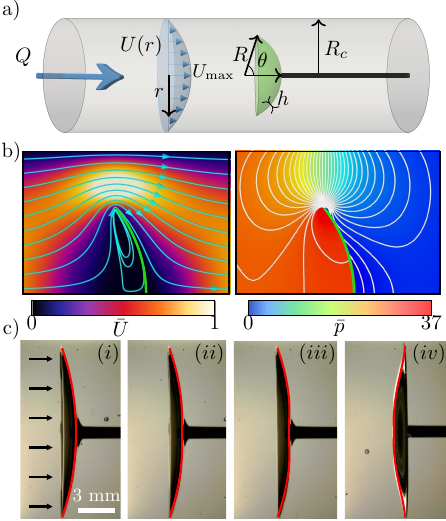}
\caption{(a) Experimental setup: an elastic shell of opening angle $\theta$ is installed at the center of a cylindrical channel, with a fully-developed parabolic Hagen-Poiseuille flow investing the elastic shell. (b) Flow around a rigid shell with $R/h = 10^4$, $\theta = 40^\circ$, $\delta = 0.6$: streamlines are superimposed on colormaps of the velocity magnitude (left) and pressure contours are superimposed on colormaps of the pressure (right), normalized by $U_{\textup{max}}$ and $\mu U_{\textup{max}} / R_c$, respectively. (c) Experimental snapshots of a shell undergoing snapping with $R=15$ mm, $h=0.23$ mm, $\theta=30^\circ$, $R_\textup{c}=12.5$ mm, $E=1.1$ MPa, $Q=100$ mL/min, overlaid with the equilibrium shapes obtained via a 1D model with the same $w$ of the snapshots. From (i) to (iv), time increases as the shell deforms slightly up to (iii), before snapping into the final shape (iv).} 
\label{fig:fig1}
\end{figure}

Fig.~\ref{fig:fig1} depicts the shell-channel system. Shells are fabricated using VPS-32, VPS-16, and VPS-8 (vinyl-polysiloxane, Zhermack) following the coating method detailed in~\cite{lee2016fabrication}, corresponding to different Young's moduli of the shell, which are $E=1.1\pm0.1$ MPa, $E=0.61\pm0.10$ MPa, and $E=0.22\pm0.05$ MPa, respectively, determined via self-buckling tests of long strips~\cite{garg2024passive}. For these materials, the Poisson ratio is $\nu\simeq0.5$, given their incompressibility~\cite{Pezzulla2018}.  The shell thickness, $h$, is measured via a custom image-processing MATLAB code
and falls within the range $h \in [0.18,0.80]\,\text{mm}$. The shells are cut to achieve the desired opening angle $\theta$ ~\cite{supp}. We only fabricate shells that exhibit bistability, that is shells that show two stable, undeformed and deformed, configurations~\cite{taffetani2018static}
, that translates into the geometrical constraint $\theta/\sqrt{h/R}>3.1$ ~\cite{Pezzulla2018}, where $R\in[12, 22.5]$ mm is the radius of the shell and $h/R\in[0.01,0.03]$ is the slenderness ratio (Fig.~\ref{fig:fig1} (a)).
Shells are then mounted at the center of cylindrical acrylic tubes with radii $R_c \in [10,21] \,\text{mm}$~\cite{supp}. 
Silicone oil (Sigma-Aldrich, density $\rho = 970 \,\text{kg m}^{-3}$, dynamic viscosity $\mu=1$ Pa$\,$s) serves as the working fluid, with flow rates $Q \in [4.3,480]\,\text{mL/min}$ controlled via a syringe pump (Harvard Apparatus PHD ULTRA Syringe Pump 70-3007). This setup imposes a parabolic Hagen-Poiseuille flow profile of maximum velocity $U_\textup{max}$. The flow rate is achieved after a short preset ramp, whose plateau value is incrementally increased until snapping is observed~\cite{garg2024passive}. {Fig.~\ref{fig:fig1} (b) shows the velocity and pressure fields as obtained via our simulations, described later. The gap between shell and channel determines a flow acceleration associated with a sudden pressure drop, with an almost constant pressure on the shell surfaces. The deformation of the shell is recorded via a digital microscope (Dino-Lite, premier), with the aid of a backlight (Edmund Optics AI Side-Fired Backlight, $2" \times 2"$, White) to enhance visualization (Fig.~\ref{fig:fig1} (c)). The Reynolds number based on the channel diameter $\textup{Re}= 2 \rho U_\textup{max}R_c/\mu$ is on average $0.2$ and at most $0.56$, ensuring negligible effects of the fluid inertia in our experiments~\cite{brenner1961oseen}. 

If the flow rate is below a critical value, the shell will return to its original configuration once the flow is reset to zero; however, above a critical flow rate, the shell will completely invert its curvature via a snapping instability and settle into a higher-energy stable configuration~\cite{misbah2016}.
The observed deformation close to snapping (Fig.~\ref{fig:fig1} (c)) is reminiscent of the prototypical example of shell instability, the pressure buckling of a hemispherical shell clamped at the equator~\cite{Zoelly1915,Hutchinson1967}. Indeed, while the boundary conditions are different between the classical case and our configuration (a clamped versus a free boundary), the loading conditions are rather similar since they are both mostly driven by pressure.
More importantly, 
both cases exhibit a limit-point instability, characterized by a subcritical bifurcation~\cite{misbah2016},
where the shell snaps into a mirror-buckled shape of either a portion of the body (classical case) or its entirety (our configuration). 
This analogy suggests that, in our case, the critical pressure that triggers a snapping instability will exhibit a scaling similar to the classical case, that is $\Delta p_b = \frac{2 E}{\sqrt{3(1-\nu^2)}}\left(\frac{h}{R}\right)^2$~\cite{Zoelly1915}. 

By means of dimensional analysis, we thus introduce the Cauchy number $C_\textup{Y} = \mu U_\textup{max}R^2/\bar{B}$, where $\bar{B}=2E h^3/\sqrt{3(1-\nu^2)}$ is the rescaled bending stiffness of the shell~\cite{garg2024passive}. The dimensionless number $C_Y$ represents the ratio between the fluid ($\sim\mu U_\textup{max}/R$) and elastic stresses ($\sim Eh^3/R^3$), thereby combining some geometrical parameters of the system with the material parameters of the shell and the fluid~\cite{Gosselin2010}. Our analogy with the pressure buckling case suggests that fluid-induced snapping should occur when the pressure loading experienced by the shell, $\Delta p$, is comparable to the classic prediction by Zoelly. If we denote the critical, snapping, pressure loading as $\Delta p_\textup{s}$, this can be expressed as $\Delta p_\textup{s} \sim \Delta p_\textup{b}$ or $C_Y^\textup{cr}  \sim R/h$, in terms of the critical Cauchy number, $C_Y^\textup{cr}$.
The remaining geometrical parameters at play are the opening angle of the shell $\theta$ and the hydrodynamic confinement ratio, defined as the ratio between the tube radius and the axially projected radius of the shell, i.e., $\delta = (R \sin \theta )/ R_c$. In our experiments, a critical flow rate thus corresponds to a critical Cauchy number $C_Y^{\textup{cr}} = F(\theta,\delta,h/R)$, indicating the fluid-induced snapping instability, with the function $F(\theta,\delta,h/R)$ being the subject of the remainder of the paper.
As a robust protocol for experiments and numerical simulations, this critical condition is identified as the minimum flow rate $Q$, or velocity $U_\textup{max}$, such that the shell snaps and does not return to its original state, if the flow is reset to zero.

\begin{figure}[t!]
\centering
\includegraphics[scale=1.05]{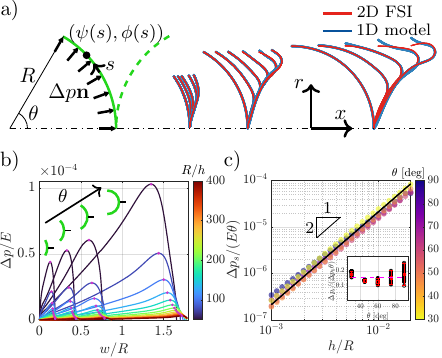}
\caption{(a) Sketch of profile curve of the shell with relevant quantities for the 1D model (left). Deformed shapes for the same axial displacement at the maximum colatitude $w$, until snapping, for fixed $R/h=50$ and $\theta=35^\circ$ (left), $\theta=60^\circ$ (center), $\theta=90^\circ$ (right): 2D FSI for fixed confinement ratio $\delta=0.7$ against 1D model. (b) Normalized equilibrium pressure $\Delta p/E$ as a function of $w$, for different $h/R$ and $\theta=30^\circ,\,45^\circ,\,60^\circ,\,90^\circ$. The maximum in the curve marks the snapping pressure. (c) Snapping pressure $\Delta p_s$ (rescaled with $E\theta$) as a function of $h/R$, for different values of $\theta$ according to the color bar. In the inset: rescaled snapping pressure (normalized by Zoelly's critical pressure and $\theta$) as a function of $\theta$.}
\label{fig:fig2}
\end{figure}

To gain a more quantitative and mechanistic understanding of the snapping instability, we develop a first-order Koiter shell model based on the Kirchhoff-De Saint Venant strain energy, valid for small strains but large displacements~\cite{Niordson1985}. As shells deform axisymmetrically until the instability (Fig.~\ref{fig:fig1} (c)), we simplify the model to be axisymmetric, and thereby a 1D model. Fig.~\ref{fig:fig2} (a) shows the undeformed geometry of the mid-surface of the shell in solid green and the deformed, snapped configuration, in dashed green. The curvilinear coordinate $s$ represents the colatitude of the shell and uniquely identifies a material point in the mid-surface, with coordinates $(\psi(s),\phi(s))$ in the $(x,r)$ coordinate system~\cite{Pezzulla2018}. The geometry of the shell is entirely described by its mid-surface
~\cite{Niordson1985,Oneill1997,Pezzulla2018}, with the unit normal vector $\mathbf{n}$, and the elastic energy of a first-order axisymmetric shell model $\overline{\mathcal{U}}_e$ is split additively into stretching and bending components~\cite{supp}. 
Thin shells, such that $h/R\ll1$, will favor bending over stretching given the different energy costs~\cite{Niordson1985}. We model the hydrodynamic force acting on the shell as a live pressure difference $\Delta p$, neglecting shear stresses exerted by the fluid, and add its energy potential $\overline{\mathcal{U}}_{\mathrm{p}}$ to the elastic energy of the shell to find the total energy of the system $\overline{\mathcal{U}} = \overline{\mathcal{U}}_e + \overline{\mathcal{U}}_{\mathrm{p}}$.
For a given displacement at $s=\theta$, $\overline{\mathcal{U}}$ is minimized to determine axisymmetric equilibrium configurations and pressure $\Delta p$ for shells clamped at the pole, with varying opening angle $\theta$ and $h/R$, via a custom weak-form based formulation in COMSOL Multiphysics~\cite{Pezzulla2018,Pezzulla2019,supp}. 
The right side of Fig.~\ref{fig:fig2} (a) depicts equilibrium shapes for three different opening angles and increasing tip displacement until snapping, for $R/h=50$, as obtained via the 1D model (solid blue) and a 2D fluid-structure model (solid red), described later in the text. Although the 1D model does not take into account neither the shear stresses acting on the shell nor the effect of confinement, we observe a very good agreement between 1D and 2D, albeit deteriorating for larger $\theta$ near the boundary. The 1D model also shows good quantitative agreement with the experimentally observed deformed shapes, as Fig.~\ref{fig:fig1} (c) shows, thus suggesting even more that the reduced-order model is adequate to describe the elasto-hydrodynamic interactions in our problem.

Fig.~\ref{fig:fig2} (b) shows the results of our parametric analyses with the 1D model in terms of $\Delta p/E$ versus $w/R$, where $w$ is the axial displacement for $s=\theta$. The non-monotonic behavior substantiates our intuition on the existence of a limit-point instability that takes place at the maximum value of pressure, which we term snapping pressure. 
We plot the snapping pressure $\Delta p_s$ as a function of $h/R$ in Fig.~\ref{fig:fig2} (c), for different values of the opening angle, and find that our results follow Zoelly's quadratic scaling $\Delta p_s\propto E(h/R)^2$, as long as pressure is rescaled with $\theta$. 
Although the prefactor slightly varies with $\theta$ within the range $[0.09,0.25]$, a fair collapse is achieved with the average prefactor 0.16, leading to
\begin{equation}
    \Delta p_s \simeq 0.16 \theta  \bar{B}\left(\frac{h}{R}\right)^2 \Rightarrow\, C_Y^\textup{cr}  \sim \theta (R/h)\,,
    \label{eq:ps}
\end{equation}
thereby enriching our scaling of the critical Cauchy number.
Notice, however, how the confinement ratio is still missing in the global picture, as we now have to proceed to the development of a hydrodynamic model.

To make progress, we perform 2D axisymmetric FSI simulations by solving the dimensionless Stokes equations coupled with the balance equations of Hookean solids, assuming small strains but large displacements, and enforcing stress continuity at the fluid-solid interface~\cite{garg2024passive,supp}. 
To derive a closed analytical solution for the critical Cauchy number, we consider a simplified framework. At snapping, despite some radial deformations, the flow experiences a pressure drop similar to that caused by a circular disk, see Fig.~\ref{fig:fig1} (b). This approximation becomes more accurate as the opening angle decreases.
We thus approximate the hydrodynamic pressure acting on the shell as the one on a flat disk with a radius equal to the projected radius, $R_p = R \sin\theta$, and negligible thickness. Van Dyke analytically derived an effective expression for the drag of a translating flat disk inside a circular tube using an asymptotic series~\cite{vanDyke1974, ZIMMERMAN20027}. We then adjust this drag expression, $D_W$, to account for a disk subject to a Hagen-Poiseuille flow within the tube, following Wakiya~\cite{wakiya1957}. The average pressure drop across the shell is eventually calculated by dividing the drag $D_W$ by the projected area $\pi R_p^2$, leading to
\begin{equation}
{\Delta p_W} = \frac{D_W}{\pi R_p^2} = \frac{16\mu U_{\textup{max}}}{\pi R_p} \left(1 - \frac{2}{3} \delta^2 \right) \frac{1}{(1 - \delta)^2} \sum_{n=0}^{\infty} d_n \delta^n\,,
\label{eq:pw}
\end{equation}
where the values of the coefficients $d_n$ and a comparison of this pressure drop formula with numerical simulations are provided in the SI \cite{supp}. By combining our prediction for the snapping pressure $\Delta p_s$ in Eq. \eqref{eq:ps}, from shell mechanics, with the pressure drop $\Delta p_W$ in Eq. \eqref{eq:pw}, from the hydrodynamic model, we derive a closed-form expression for the critical Cauchy number
\begin{equation}
    C_Y^{\textup{cr}} \simeq 0.1{ \pi} \theta \sin(\theta) \Gamma(\delta)^{-1} \left(\frac{R}{h}\right):= F(\theta, \delta, h/R)\,,
\end{equation}
where the dependence on all parameters is now explicit. The function $\Gamma(\delta) := \left( 1 - \frac{2}{3} \delta^2 \right) \frac{1}{(1 - \delta)^2} \sum_{n=0}^{\infty} d_n \delta^n$ captures the geometrical effects induced by hydrodynamic confinement. 
\begin{figure}[ht]
\centering
\includegraphics[scale=0.9]{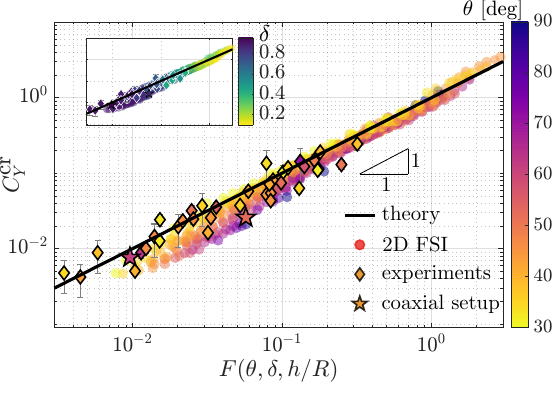}
\caption{Critical Cauchy number as a function of $F(\theta, \delta, h/R)$: theoretical prediction (black solid line), numerics (colored circles) and experiments (colored diamonds with error bars). The colormap highlights the value of $\theta$ in each experiment and simulation. In the inset, the same plot is reported, but the colormap depicts the confinement ratio. Stars represent critical experimental Cauchy values for the snapping valve in the coaxial channel setup.}
\label{fig:fig4}
\end{figure}
Fig.~\ref{fig:fig4} summarizes our results in terms of the critical Cauchy number $C_Y^{\textup{cr}}$ as a function of $F(\theta, \delta, h/R)$.
The agreement among theoretical (solid line), numerical (colored circles), and experimental (diamonds) results is remarkable, since no fitting parameters have been employed \cite{supp}.
The data spans nearly three orders of magnitude on both axes, covering a wide range of confinement ratios ($0.1<\delta<0.95$) and opening angles ($25^\circ<\theta <90^\circ$), thereby supporting a robust scaling law for the snapping of spherical shells in viscous flow.

We now demonstrate how bistable shells can be used to design a passive valve that can isolate vessels and abruptly change the flow-pressure relationship in a fluidic channel.
To this aim, we 3D printed additional setups of co-axial fluidic channels, with the ratio between the outer radius of the internal tube $R_i$ and the inner radius of the external one $R_c$, $\xi=R_i/R_c$, equal to $0.76$ and $0.91$, respectively, and with $R_c=10,\, 25$ mm, and the length of the internal channel equal to half of the length of the outer channel (Formlabs, Form3 printer, Clear resin). We then placed a spherical shell at the inlet of the internal channel as shown in Fig.~\ref{fig:fig5}, and connected the channel to the syringe pump used so far in the paper. 
Both shells for the two setups are characterized by $E=1.1\pm0.1$ MPa and $R=22.5$ mm, with $h=267 \pm 9$ $\mu$m and $\theta=53^\circ$ for $\xi=0.76$, and with $h=273\pm 12$ $\mu$m and $\theta=63^\circ$ for $\xi=0.91$. 
\begin{figure}[b]
\vspace{0.2cm}
\centering
\includegraphics[scale=1.07]{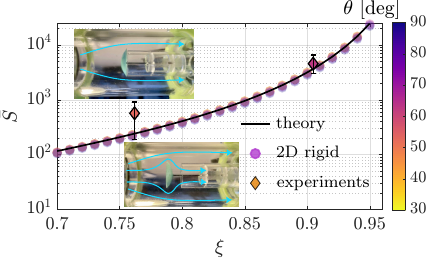}
\vspace{-10pt}
\caption{Difference in pressure drop per unit length before and after snapping as a function of the ratio between inner and outer tube radii, nondimensionalized by $\mu U_\textup{max}/R_c^2$: theory (black line), numerics with rigid shells (colored dots) and experiments (diamonds with error bars). Colors represent different values of $\theta$, as indicated by the colorbar. The insets are snapshots of the shell before (bottom) and after (top) snapping, effectively sealing the inner tube.} 
\label{fig:fig5}
\end{figure}
The insets in Fig.~\ref{fig:fig5} shows the two states of our system: at the bottom, the shell is in its reference configuration, thereby allowing the fluid to flow in both the inner and outer channels; at the top, the shell has snapped, preventing the fluid to flow within the internal channel. The experimental critical Cauchy number follows our prediction despite the internal channel placed downstream, as shown by the two stars in Fig.~\ref{fig:fig4}, corresponding to the experimental runs with these coaxial setups. To measure the pressure drop across the inner channel, we placed a $1$ bar relative pressure sensor (Omega Engineering, PXM409-001BGUSBH) just upstream of the channel. This allowed us to quantify the relative increase in pressure drop per unit length, defined as $S=(\Delta p_a-\Delta p_b)/L$, where $\Delta p_a$ and $\Delta p_b$ are the (dimensionless) pressure drops measured after and before snapping, respectively, over the length $L$ of the inner tube. All pressures are non-dimensionalized by $\mu U_\textup{max}/R_c$. 
Before snapping, the total flow rate $Q$ is the sum of the flow through the inner tube and the surrounding annulus, each described by a Poiseuille flow. After snapping, the flow is confined entirely to the annular region, which now accommodates the full flow rate $Q$. Applying classical expressions for cylindrical and annular Poiseuille flow, we derive \cite{supp}
\begin{equation}
S= \frac{8\mu Q}{\pi R_c^4}\!\!\left[\! \left({1-\frac{\left(1-\xi^2\right)^2}{\text{ln}\left(\xi^{-1}\right)}}\right)^{-1}\!\!\!\!\!\!\!\!-\left({1-\xi^4-\frac{\left(1-\xi^2\right)^2}{\text{ln}\left(\xi^{-1}\right)}}\right)^{-1}\!\right].
\label{eq:S}
\end{equation}
Fig.~\ref{fig:fig5} shows the nondimensional pressure drop difference between the two states, $\bar{S}=S/(\mu U_\textup{max}/R_c^2)$, as a function of $\xi$, as obtained via our theory (Eq. \eqref{eq:S}, solid line), 2D simulations with rigid shells in the initial and inverted curvature configurations (circles), and experiments (diamonds). 
We find good agreement across our different methods, thus conveying both the feasibility of our snapping valve and the quality of our theoretical and numerical frameworks, across order of magnitudes.

In summary, we demonstrated that a bistable spherical shell can be used to devise a passive valve for nonlinear hydraulic control at low Reynolds numbers. Our combined experimental, numerical, and theoretical analysis revealed that the shell exhibits a sharp, reversible transition triggered by a critical flow rate. Although our experiments are at the millimeter scale, our conclusions are rooted in dimensional analysis, leading to results applicable across different length scales. Our model applies in the viscous regime: simulations of the Navier–Stokes equations for confined thin disks show a progressive deviation that becomes relevant for $\textup{Re}>2$, a safe threshold for the validity of our theory \cite{supp}. The theory further assumes small strains, typically satisfied by thin shells. While our model is limited to 2D axisymmetric cases, it can serve as a reliable first-order approximation for small deviations from axisymmetry.
Finally, we speculate that fluidic networks can be equipped with a variety of snapping valves, including those exhibiting more than two stable states, to achieve decentralized control and a form of mechano-fluidic, or physical, intelligence~\cite{Sitti2021}. Indeed, our valve exhibits memory~\cite{Preston2019}, and multiple valves can interact within a network to offer sensing and self-regulation without any electronic components—a perspective we hope will motivate further studies within the nascent field of soft hydraulics~\cite{Wehner2016,Christov2021}.

\begin{acknowledgments}
This work was supported by a research grant (VIL50135) from Villum Foundation. M.P. acknowledges also the support from the Thomas B. Thriges Fond. M.P. conceived the project and supervised the research with input from P.G.L. P.G.L. and M.P. wrote the manuscript. All authors revised the manuscript. P.G.L. and M.P. developed the theoretical models. M.P. prepared the FSI numerical setup, and H.G. performed the FSI simulations. V.\O{}-C., L.K.R., and A.M. designed and built the experimental setup. V.\O{}-C. and L.K.R. conducted the experiments with input from H.G.
\end{acknowledgments}

\emph{Data availability} -- The data that support the findings of this article are openly available \cite{data}.

\balancecolsandclearpage
\newpage

\onecolumngrid

\begin{center}
\vspace{0.2cm}\noindent \textbf{\large Supporting information}
\end{center}

\setcounter{equation}{0}
\setcounter{figure}{0}
\setcounter{table}{0}
\setcounter{page}{1}
\makeatletter
\renewcommand{\theequation}{S\arabic{equation}}
\renewcommand{\thefigure}{S\arabic{figure}}

\section{Experimental apparatus}

A schematic of the complete experimental setup is shown in Figure \ref{fig:setup} (left). The channel is placed on two aluminum rails, with a backlight  (Edmund Optics AI Side-Fired Backlight, $2" \times 2"$, White) placed below the channel to help with image processing. The digital microscope (Dino-Lite, premier) is mounted at the top and used to record and capture the shell deformation. A family of channels with different radii is used to test different geometries. The shell, has shown also in the inset in the top-left corner, is placed within the channel, held by a carbon rod. To make sure the carbon rod is perfectly centered, a 3D printed support is placed within the channel to host the carbon rod. The syringe pump (Harvard Apparatus PHD ULTRA Syringe Pump 70-3007) drives the flow within the system, with silicone oil (Sigma-Aldrich, kinematic viscosity, $\eta=1000$ cSt and density $\rho= 970$ kg m$^{-3}$) as the working fluid. The syringes are ACONDE 150 mL plastic syringes. Figure \ref{fig:setup} (right) shows the 3D printed setup, designed to hold the carbon rod in contact with the north pole of the metal sphere, orthogonal to the tangent plane at that point. In the picture, the VPS polymer is crosslinking.

\begin{figure}[h]

\includegraphics[scale=0.841]{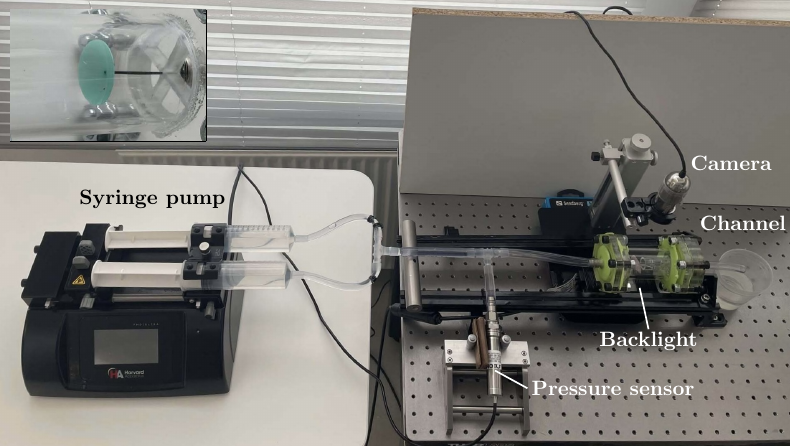}
    \includegraphics[scale=0.44]{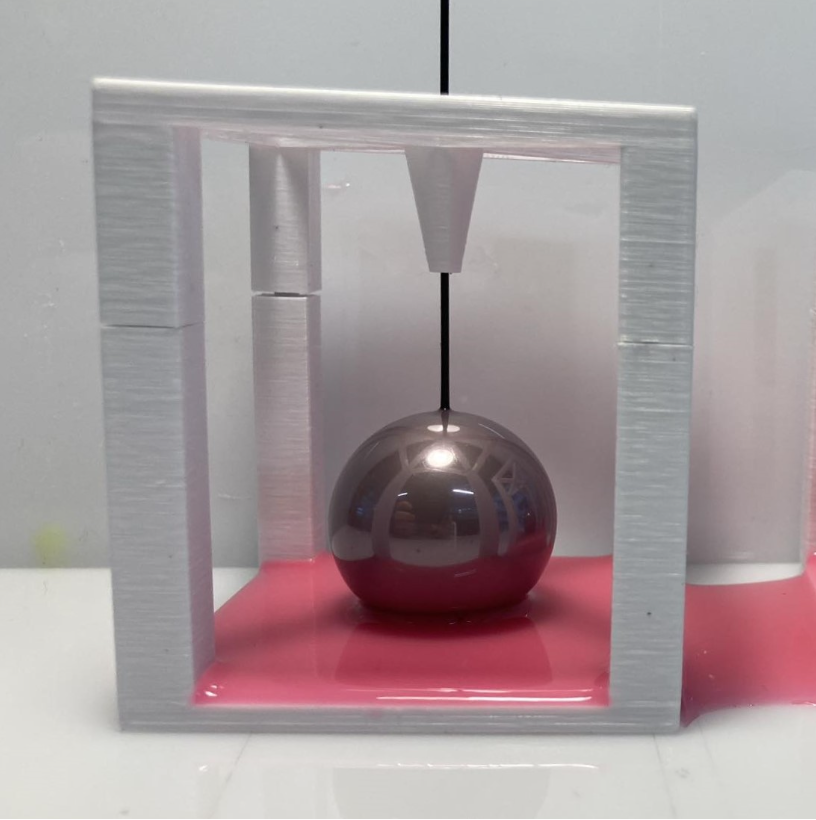}

\caption{Photo of the experimental apparatus comprising the syringe pump, the pressure sensor, the scientific camera with a backlight, and the cylindrical channel. A detail of the VPS shell, the carbon rod, and its support structure are depicted in the top left corner (left). Picture of the 3D printed setup that holds the carbon rod in contact with the north pole of the metal sphere: the VPS polymer is crosslinking (right).}
\label{fig:setup}
\end{figure}

\vspace{0.2cm}\noindent \textbf{Shell fabrication.} We fabricate shells by following the coating technique presented in ~\cite{lee2016fabrication}. We prepare the polymeric mixture by mixing the bulk polymer with the curing agent at a $1:1$ ratio in mass. The solution is mixed in a centrifugal mixer (Thinky Mixer ARE-250CE) for $20$ s at $2000$ rpm. Then, the solution is poured on rigid metal spheres (radii $R\in[12, 22.5]$ mm), while a carbon rod ($1$ mm in diameter) is held in contact with the north pole of ball via a 3D printed setup (see Fig. S1 in the SI). The solution thermally crosslinks into an elastomeric shell at room temperature, with the carbon rod naturally glued to the shell at its north pole. As detailed in the main text, shells with specific half-angles $\theta$ are cut by first placing an acrylic sheet with a circular hole on top of the sphere, and then cutting around the parallel with a scalpel.
These shells are fabricated using VPS-32, VPS-16, and VPS-8 (vinyl-polysiloxane, Zhermack) following the coating method detailed in~\cite{lee2016fabrication}, corresponding to different Young's moduli of the shell, which are $E=1.1\pm0.1$ MPa, $E=0.61\pm0.10$ MPa, and $E=0.22\pm0.05$ MPa, respectively. These values are determined via self-buckling tests of long strips~\cite{garg2024passive}. For these materials, the Poisson ratio is $\nu\simeq0.5$, given their incompressibility~\cite{Pezzulla2018}. 
The shell thickness, $h$, is measured via a custom image-processing MATLAB code applied on photographs taken by a digital microscope (Dino-Lite, premier), and falls within the range $h \in [0.18,0.80]\,\text{mm}$. 

\vspace{0.2cm}\noindent \textbf{Hydraulic setup.} The cylindrical channel (Fig. 1) is placed on two aluminum rails, with a backlight  (Edmund Optics AI Side-Fired Backlight, $2" \times 2"$, White) placed below the channel to help with image processing. The digital microscope (Dino-Lite, premier) is mounted at the top and used to record and capture the shell deformation. A family of channels with different radii $R_c \in [10,21] \,\text{mm}$ is used to test different geometries. The shell, has shown also in the inset in the top-left corner, is placed within the channel, held by a carbon rod. To make sure the carbon rod is perfectly centered, a 3D printed support is placed within the channel to host the carbon rod. The syringe pump (Harvard Apparatus PHD ULTRA Syringe Pump 70-3007) drives the flow of silicone oil (Sigma-Aldrich, kinematic viscosity, $\eta=1000$ cSt and density $\rho= 970$ kg m$^{-3}$) within the system, with flow rates $Q \in [4.3,480]\,\text{mL/min}$. The syringes are ACONDE 150 mL plastic syringes. A comprehensive picture of the entire experimental setup can be found in the SI.
This setup imposes a parabolic Hagen-Poiseuille flow profile of maximum velocity $U_\textup{max}$. In a typical experiment, the flow rate is achieved after a short preset ramp, whose plateau value is incrementally increased until snapping is observed~\cite{garg2024passive}. The Reynolds number based on the channel diameter $\textup{Re}= 2 \rho U_\textup{max}R_c/\mu$ is on average $0.39$ and at most $1.12$, ensuring negligible effects of the fluid inertia in our experiments~\cite{brenner1961oseen}.

\section{Uncertainty analysis of experimental results}

The error propagation on the Cauchy number \( C_{\textup{Y}} = \frac{\mu U_{\textup{max}} R^2}{\bar{B}} \) stemming from experimental uncertainties in \( \mu \), \( U_{\textup{max}} \), \( R \), \( E \), and \( h \) is estimated by considering the sensitivities of the Cauchy number to each of these parameters:
\begin{equation} \label{partuncertainity}
\begin{minipage}{\textwidth}
\begin{equation} \notag
\frac{\partial C_{\textup{Y}}}{\partial \mu} = \frac{U_{\text{max}} R^2}{\bar{B}}, \quad
\frac{\partial C_{\textup{Y}}}{\partial U_{\text{max}}} = \frac{\mu R^2}{\bar{B}}, \quad
\frac{\partial C_{\textup{Y}}}{\partial R} = \frac{2\mu U_{\text{max}} R}{\bar{B}}, \quad
\frac{\partial C_{\textup{Y}}}{\partial E} = -\frac{\mu U_{\text{max}} R^2}{E \bar{B}}, \quad
\frac{\partial C_{\textup{Y}}}{\partial h} = -\frac{3\mu U_{\text{max}} R^2}{h \bar{B}}\,.
\end{equation}
\end{minipage}  
\end{equation}

The overall uncertainty on the experimental Cauchy number (i.e., the vertical error bar in Fig.~4 of the main text) thus reads:
\begin{equation} \label{uncertainity}
\delta C_{\textup{Y}} = \sqrt{\left(\frac{\partial C_{\textup{Y}}}{\partial \mu} \delta \mu\right)^2 + \left(\frac{\partial C_{\textup{Y}}}{\partial U_{\textup{max}}} \delta U_{\textup{max}}\right)^2 + \left(\frac{\partial C_{\textup{Y}}}{\partial R} \delta R\right)^2 + \left(\frac{\partial C_{\textup{Y}}}{\partial E} \delta E\right)^2 + \left(\frac{\partial C_{\textup{Y}}}{\partial h} \delta h\right)^2}.
\end{equation}

In our case, we noticed a high sensitivity of the Cauchy number with respect to the thickness and radius uncertainties. 
As concerns the shell radius \( R \), since the shells are fabricated using metal balls of negligible tolerance, the radius is highly reproducible, but the thickness varies by about 8\% from the pole to the equator, for $\theta=90^\circ$, according to measurements and in agreement with \cite{lee2016fabrication}. Thickness was measured using a digital microscope, and the error, based on the camera resolution, is about 2\%, dependent on the considered shell, to be added to the previous uncertainty. Similarly, the error estimate for the Young's modulus is kept below 5\%, determined through statistical measures of successive tests with different samples. The error in viscosity is estimated from the datasheet provided by the manufacturer, where its variation with temperature is reported. Our laboratory can experience a temperature variation of up to $\pm1^\circ$C, leading to a relative error in viscosity of up to 5\%, in agreement with \cite{garg2024passive}. The error estimation for the maximum velocity \( U_{\textup{max}} \) is determined for each shell experiment from the experimental quantization in the flow rate when finding the threshold. These variations range from \( 10^{-5}\) to \(10^{-4}\)~m/s, specific to each experiment.

\subsection{Other sources of experimental scatter}

As concerns manufacturing, uneven cutting edges can be present since the shells are cut out with a scalpel. The uneven cutting edges slightly change the local opening angle of the shell. Using simple geometry, the error in the opening angle during the cut can be reasonably estimated as the ratio between the scalpel blade thickness (0.3~mm) and the shell radius. In the worst case, this gives an error of about $\Delta \theta \sim 2.5\%$, which could in principle be included in the $x$-axis of Fig.~4. However, the corresponding error bar would be smaller than the marker size. The rod is fixed to the shell during the coating procedure, resulting in a larger support for the shell compared to theory. This point has been addressed and discussed in the next section, showing variations of the order of a few percent in the critical pressure, and thus neglected. The asymmetry of the carbon rod attachment point with respect to the shell can be another source of scatter. Asymmetry can lead to skewing of the shell during the experiment, potentially compromising its integrity, which was the reason for the introduction of a rod aligner (see Fig.~\ref{fig:setup}) that significantly reduced possible misalignments of the shell with respect to the channel. Finally, during experiments, swelling of silicone oil within VPS progressively occurs, leading to a steady-state linear stretching factor of 5\%. This results in an increase in shell radius and thickness, and in modifications of the Young's modulus. To reduce this effect and minimize this source of uncertainty, the reported experiments were performed only with new shells, immediately after their positioning in the channel.

\vspace{0.6cm}

\section{Differential geometry framework and 1D model} 
The 1D model is based on a shell model originally developed for non-Euclidean shells~\cite{Pezzulla2018} and later extended to classical shell mechanics problems with pressure loading~\cite{Pezzulla2019}. The energy, written in the main text, is split into stretching and bending components and is valid for small strains and large displacement gradients. The key ingredients of the energy are the first and second fundamental forms of the mid-surface, reduced to a profile curve because of axisymmetry.

We refer the reader to the cited works~\cite{Pezzulla2018,Pezzulla2019}, while providing here some basic notions from differential geometry and some details on the numerical implementation in COMSOL Multiphysics. The profile curve of the shell is a real segment $s\in[0, \theta]$, with $\theta=0$ corresponding to the north pole. Within this setup, the first and second fundamental forms $\mathring{\mathbf{a}}$ and $\mathring{\mathbf{b}}$ of the undeformed mid-surface are known, while the ones of the deformed mid-surface $\mathbf{a}$ and $\mathbf{b}$ can be expressed as a function of the placement vector $(\psi(s),\phi(s))$~\cite{Oneill1997,Pezzulla2018,Pezzulla2019}.

The elastic energy of a first-order axisymmetric shell model, non-dimensionalized by $\pi E h R^2 /\left(4\left(1-\nu^2\right)\right)$, can be written as
\begin{equation}
\overline{\mathcal{U}}_e = \int_0^\theta\left[(1-\nu) \operatorname{tr}\left(\mathbf{a}-\stackrel{\circ}{\mathbf{a}}\right)^2+\nu \operatorname{tr}^2(\mathbf{a}-\stackrel{\circ}{\mathbf{a}})\right] \mathrm{d} s 
+\frac{h^2}{3} \int_0^\theta\left[(1-\nu) \operatorname{tr}(\mathbf{b}-\stackrel{\circ}{\mathbf{b}})^2+\nu \operatorname{tr}^2(\mathbf{b}-\stackrel{\circ}{\mathbf{b}})\right] \mathrm{d} s\,,
\end{equation}
where $\operatorname{tr}$ denotes the trace operator in the coordinate system defined by $\mathring{\mathbf{a}}$. The first term of the elastic energy represents the stretching energy, penalizing changes in lengths and angles, while the second term represents the bending energy, penalizing changes in curvature~\cite{Niordson1985}. We model the hydrodynamic force acting on the shell as a live pressure difference $\Delta p$, neglecting shear stresses exerted by the fluid, via its (dimensionless) energy potential~\cite{Pezzulla2019}:
\begin{equation}
        \overline{\mathcal{U}}_{\mathrm{p}} =
         \frac{8\left(1-\nu^2\right)}{3 h R^2} \frac{\Delta p}{E}\int_0^{\theta} \phi(\psi\phi_s-\psi_s\phi) \mathrm{~d} s\,,
\end{equation}
where $()_s$ denotes the first derivative with respect to $s$.

The total dimensionless energy, including the elastic component and the potential of the live pressure, is then minimized in COMSOL Multiphyisics (v6.2) within the Weak Form PDE interface, with the addition of symmetric boundary conditions at the north pole and free boundary conditions for $s=\theta$~\cite{Pezzulla2019}. As the carbon rod that holds the shell in place has a finite radius, we performed a parametric analysis to assess the influence of the size of the clamped area on our results. To this aim, we varied the angular size of the clamped area from $0^\circ$ to $\theta/20$, finding a relative change of the critical pressure of up to $3\%$.

\section{FSI simulations}
Fluid-structure interactions simulations, for both rigid and flexible shells, are carried out in COMSOL Multiphysics (v6.2), within the Fluid-Solid interface, for 2D axisymmetric problems. We solve the Stokes equations, paired with the balance equations for linear elastic solids, in dimensionless form. A time-dependent solver is employed to solve for the shell deformation in response to an imposed inlet parabolic Poiseuille flow profile, ramping from zero to steady state. No-slip conditions are assigned on the lateral side of the channel and the boundary of the shell; a condition of zero pressure is assigned on the outlet. In this time-dependent setting, the threshold is identified as the minimum value of inlet velocity for which the shell snaps. Quadratic Lagrangian shape functions are used for the velocity field of the fluid and the displacement field of the solid, whereas linear shape functions were used for the pressure field within the fluid. A convergence study is also performed: the model is considered at convergence if further mesh refinement corresponds to a relative variation of the critical inlet velocity smaller than $1\%$.

Simulations of the rigid case are carried out within the same framework, omitting the FSI interface while retaining the creeping flow solver. To neglect thickness effects, a numerically ideal zero thickness is approximated by assigning a value of $10^{-4}$, in units of the tube radius, for both disks and shells. In the coaxial tube setup, the same numerical thickness of $10^{-4}$ is also applied to the inner tube. The projected radius of the shell $R_p$ is set equal to the inner tube radius $R_i$, and its distance from the inner tube is chosen such that, once the shell curvature is inverted (mimicking the snapped state), it just contacts the inner tube boundary, effectively creating a single rigid surface that isolates the inner tube. To match experimental conditions, the inner tube terminates near a single outlet where zero pressure is imposed (this corresponds to atmospheric pressure in the experiment, achieved slightly downstream of the coaxial tube setup).

\vspace{0.2cm}\noindent \textbf{Pressure drop.} We measure the average pressure drop ${\Delta p}$, non-dimensionalized by $\mu U_\textup{max}/R_c$, for rigid disks of zero thickness, and plot it against ${\Delta p_W}$ in Fig.~\ref{fig:sim_SI}.
\begin{figure}[h]

\includegraphics[scale=1.2]{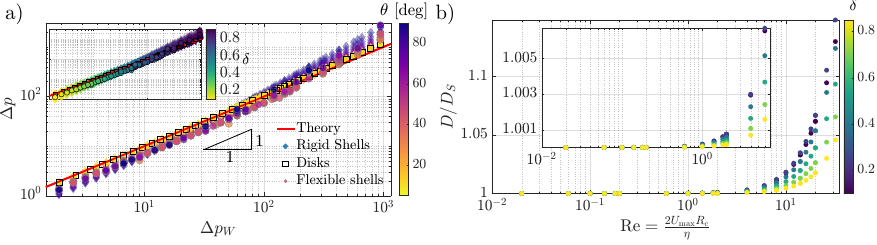}

\caption{a) Measured average pressure drop for rigid disks (squares), rigid spherical shells of negligible thickness, $R/h = 10^4$ (colored diamonds), and elastic shells at the snapping threshold (colored dots) for various $R/h$, plotted against the theoretical pressure drop from Eq.~4 in the main text (red solid line). Colors represent different values of $\theta$, as indicated by the colorbar. In the inset: same plot, but with the colormap indicating the confinement ratio. b) Flow around a thin disk: ratio between the drag for nonzero Reynolds number and drag of the Stokes problem, as a function of the Reynolds number. The colormap identifies different levels of disk confinement.}
\label{fig:sim_SI}
\end{figure}
The theoretical line denotes $\Delta p=\Delta p_W$ and agrees well with the pressure drops for all simulated cases (squares). Notice that the agreement is already satisfying with only four coefficients in the asymptotic expansion in Eq. (2) in the main text, and no significant changes are observed with additional terms. To test our model further, we measure the pressure drop for rigid shells of zero thickness (colored diamonds), finding good agreement with our theory. In this case, the drag is evaluated by considering only the pressure drop projected along the normal direction, neglecting viscous stresses. Finally, we also plot the measured average pressure drop at snapping for 2D FSI simulations of flexible shells (colored circles), for varying values of the opening angle, $h/R$, and confinement ratio $\delta$. These values closely match our theoretical prediction, thus validating our simple, yet efficient analytical expression. 
In the inset, the same plot is shown, this time with the color coding representing the confinement ratio. Remarkably, our model captures the pressure drop across three orders of magnitude, with only minor deviations at the extreme ends.

For unconfined thin disks and Reynolds numbers ($UR/\eta$) below unity, inertial effects are commonly assumed to be neglected. Since in our case we have confinement effects, we performed a systematic study. In COMSOL, we thus solved the incompressible Navier-Stokes equations for a thin disk, where we imposed different Reynolds numbers ($2UR_c/\eta$) in the range $0.01 < \textup{Re} < 32$, for confinement $0.1 < \delta < 0.85$. We compared the drag ($D \approx \int_S \Delta p$) with the drag computed using the Stokes solver ($D_s$) to assess the extent to which the viscous assumption for the equivalent disk remains valid. As shown in Fig. \ref{fig:sim_SI} (b), the ratio $D/D_s$ starts to deviate from unity and depends on the confinement ratio at a Reynolds number, based on the tube diameter, of about $2$. For large values of Reynolds numbers, we observe a clear blow-up. We can thus conclude that for $\textup{Re}<2$ the error remains below 0.1$\%$, and this value can be assumed as a safe threshold for our theoretical models. All experimental data fall within this limit, confirming that the experiments are conducted in the viscous regime.

\section{Van Dyke coefficients}
The drag expression provided in the main text reads:
\begin{equation}
{D_W} = {16\mu U_{\textup{max}}R_p}\left(1 - \frac{2}{3} \delta^2 \right) \frac{1}{(1 - \delta)^2} \sum_{n=0}^{\infty} d_n \delta^n\,,
\end{equation}
with $d_n$ coefficients provided by an asymptotic expansion by Van Dyke \cite{vanDyke1974}. The first six coefficients are:
$
d_0=1, \
d_1=-2.137\times10^{-1},\
d_2=6.183\times10^{-1}, \
d_3=-2.357\times10^{-2}, \
d_4=1.989\times10^{-1}, \
d_5=-1.000\times10^{-1}.
$

\section{Pressure drop theory for coaxial tubes with snapping valves}

\begin{figure}[h]

\includegraphics[scale=1.05]{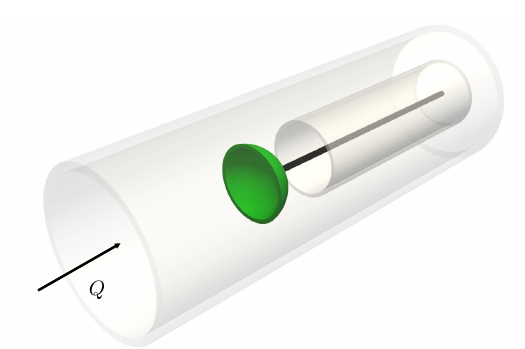}

\caption{Schematics of the coaxial fluidic channels with the snapping valve positioned at the inlet of the inner channel.}
\label{fig:coaxial}
\end{figure}

Here, we give more detail on the hydraulic model to evaluate the pressure drop increase, $S$, in the coaxial tube system, whose schematics is shown in Figure \ref{fig:coaxial}. Before snapping, the pressure drops in the inner tube, $\Delta p_i$, and in annulus between the inner and outer tubes, $\Delta p_o$, have to be equal to each other. We define them as $\Delta p_b=\Delta p_i=\Delta p_o$, where the subscript $b$ stands for before.
In addition, the inner flow rate in the inner tube is
$$Q_{i}=({\pi \Delta p_i R_i^4})/({8 \mu L})\,,$$
while the outer flow rate in the annular tube is $$Q_{o}=\frac{\pi \Delta p_o}{8 \mu L}\left(R_c^4-R_i^4-\frac{\left(R_c^2-R_i^2\right)^2}{\ln (R_c / R_i)}\right)\,.$$
Their sum must be equal to the total flow rate $Q_b=Q_i+Q_o$.
After snapping, the pressure drop given by the Poiseuille flow in the outer annulus is denoted as $\Delta p_a$, with a corresponding flow rate $Q_a$ given by
$$Q_a=\frac{\pi \Delta p_a}{8 \mu L}\left(R_c^4-R_i^4-\frac{\left(R_c^2-R_i^2\right)^2}{\ln (R_c / R_i)}\right)\,.$$
Assuming $Q_b \simeq Q_a := Q$, the pressure drop increase is defined as
$$S=\frac{1}{L}\left(\Delta p_a-\Delta p_b\right)\,,$$
leading to Eq. (4) of the main text.

To determine the pressure drops, we conducted experiments in two distinct channel configurations and recorded the time series of pressure data before and after snapping. To mitigate the effects of oscillations from the syringe pump, we analyzed the steady-state portion of the signal and extracted average, maximum, and minimum pressure values. We also removed any baseline offset by subtracting the lowest pressure value recorded before the start of the experiment. The difference in pressure drop before and after snapping was used to compute the mean and uncertainty of $S$.

\setcounter{figure}{0}
\makeatletter
\renewcommand{\figurename}{MOV.}

\newpage

\section{Supplementary movie}

\begin{figure}[!h]
    \centering
    \includegraphics[width=0.4\textwidth]{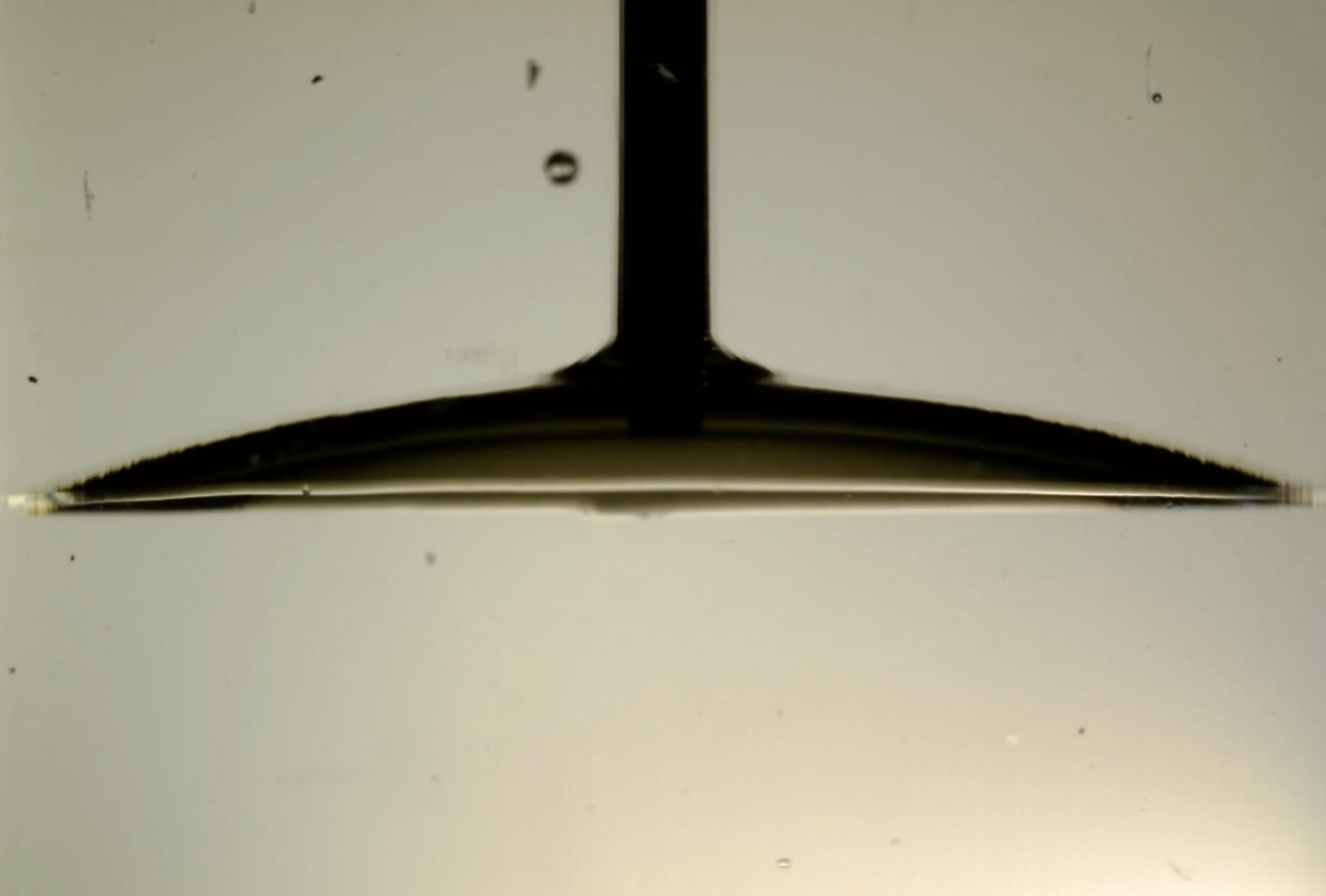}
    \caption{Experiment with $R=15$ mm, $h=0.23$ mm, $\theta=30^\circ$, $R_\textup{c}=12.5$ mm, $E=1.1$ MPa, and $Q=100$ mL/min, higher than the snapping threshold. The shell and the carbon rod can be seen in black.}
    \label{fig:shell10}
\end{figure}

\end{document}